\RequirePackage{lineno}

\documentclass[%
 amsmath,amssymb,
 prc,
nofootinbib
]{revtex4}
\usepackage{hyperref}
\usepackage{graphicx}
\usepackage{lipsum}
\usepackage{sidecap}
\usepackage[usenames,dvipsnames]{color}

\usepackage{graphicx}% Include figure files
\usepackage{bm}% bold math

\newcommand{\pt}{\ensuremath{p_{\rm{T}}}~}

\newcommand{\sqrtsn}{\ensuremath{\sqrt{s_{\rm{NN}}}}~}
\newcommand{\AuAu}{Au$+$Au~}

\newcommand{\pp}{$p+p$~}
\newcommand{\ppAA}{$p+p~\oplus$ Au$+$Au~}
\newcommand{\pAu}{$p+$Au~}
\newcommand{\dAu}{$d+$Au~}

\newcommand{\Aj}{\ensuremath{A_{\rm{J}}}}
\newcommand{\zg}{\ensuremath{z_{\rm{g}}}}
\newcommand{\rg}{\ensuremath{R_{\rm{g}}}}
\newcommand{\zsj}{\ensuremath{z_{\rm{SJ}}}}
\newcommand{\tsj}{\ensuremath{\theta_{\rm{SJ}}}}

\newcommand{\red}[1] {
  {\color{red} #1}}

\begin{document}

\preprint{\red{v13}}
 
\title{Differential measurements of jet substructure and partonic energy loss in \AuAu collisions at \sqrtsn $=200$ GeV}

\author{
M.~S.~Abdallah$^{5}$,
B.~E.~Aboona$^{55}$,
J.~Adam$^{6}$,
L.~Adamczyk$^{2}$,
J.~R.~Adams$^{39}$,
J.~K.~Adkins$^{30}$,
G.~Agakishiev$^{28}$,
I.~Aggarwal$^{41}$,
M.~M.~Aggarwal$^{41}$,
Z.~Ahammed$^{60}$
% ...
}

\address{\rm{(STAR Collaboration)}}

% ...
\address{$^{2}$AGH University of Science and Technology, FPACS, Cracow 30-059, Poland}
% ...
\address{$^{5}$American University of Cairo, New Cairo 11835, New Cairo, Egypt}
\address{$^{6}$Brookhaven National Laboratory, Upton, New York 11973}
% ...
\address{$^{28}$Joint Institute for Nuclear Research, Dubna 141 980, Russia}
% ...
\address{$^{30}$University of Kentucky, Lexington, Kentucky 40506-0055}
% ...
\address{$^{39}$Ohio State University, Columbus, Ohio 43210}
% ...
\address{$^{41}$Panjab University, Chandigarh 160014, India}
% ...
\address{$^{55}$Texas A\&M University, College Station, Texas 77843}
% ...
\address{$^{60}$Variable Energy Cyclotron Centre, Kolkata 700064, India}
% ...

\date{\today}

\begin{abstract}

The STAR collaboration presents jet substructure measurements related to both the momentum fraction and the opening angle within jets in \pp and \AuAu collisions at \sqrtsn $= 200$ GeV. The substructure observables include SoftDrop groomed momentum fraction (\zg), groomed jet radius (\rg), and subjet momentum fraction (\zsj) and opening angle (\tsj). The latter observable is introduced for the first time. Fully corrected subjet measurements are presented for \pp collisions and are compared to leading order Monte Carlo models. The subjet \tsj~distributions reflect the jets leading opening angle and are utilized as a proxy for the resolution scale of the medium in \AuAu collisions. We compare data from \AuAu collisions to those from \pp which are embedded in minimum-bias \AuAu events in order to include the effects of detector smearing and the heavy-ion collision underlying event. The subjet observables are shown to be more robust to the background than \zg~and \rg. 

We observe no significant modifications of the subjet observables within the two highest-energy, back-to-back jets, resulting in a distribution of opening angles and the splittings that are vacuum-like. We also report measurements of the differential di-jet momentum imbalance ($A_{\rm{J}}$) for jets of varying \tsj. We find no qualitative differences in energy loss signatures for varying angular scales in the range $0.1 < $ \tsj $ < 0.3$, leading to the possible interpretation that energy loss in this population of high momentum di-jet pairs, is due to soft medium-induced gluon radiation from a single color-charge as it traverses the medium.

\end{abstract}

\maketitle

\section{Introduction}

Since the start of heavy-ion collisions at RHIC, all the experiments have aimed at elucidating the properties 
of the hot and dense emergent state of matter called the quark-gluon plasma (QGP)~\cite{Adams:2005dq, 
Adcox:2004mh, Back:2004je, Arsene:2004fa}. Measurements in heavy-ion collisions aim at studying various aspects of the QGP  connected to its initial state, bulk evolution, and interactions with hard-scattered partons 
(quarks and gluons). 
Jets~\cite{Sterman:1977wj}, which are clusters of final-state particles resulting from the Quantum 
Chromodynamics (QCD) evolution, {\textit i.e.}, fragmentation of 
hard-scattered partons, are a well established probe of the QGP. The progenitor partons are produced almost 
immediately with large enough $Q^{2}$, the 4-momentum transfer squared of the hard process, 
such that its formation 
time is less than that of the QGP. For each hard-scattered parton, the resulting parton shower traverses the QGP, probing 
its entire lifetime, 
and are measured as collections of collimated final state particles (jets) in the detectors. Therefore any 
modifications to the jet energy and structure in \AuAu relative to those in \pp or \pAu
collisions~\cite{Adams:2003im, Adler:2002tq} are interpreted as effects arising due to the transport 
properties of the QGP\footnote{While cold nuclear matter effects on jet production have been recently quantified in theoretical frameworks~\cite{Kang:2015mta}, their effects on jet fragmentation are still considered to be negligible~\cite{ATLAS:2017pgl}.} and are often referred to as jet quenching~\cite{Gyulassy:1993hr, Baier:1994bd, Baier:1996sk, Zakharov:1996fv}.
First generation measurements at RHIC, which studied the phenomenon of 
jet quenching utilized high momentum hadrons as proxies for jets and found a marked suppression of high 
transverse momentum ($p_{\mathrm{T}}$) hadron yield~\cite{Adams:2003kv, Adcox:2001jp}.
In addition, in high multiplicity or central \AuAu collisions the measured yields of associated hadrons in the back-to-back azimuthal region, with respect to a high-$p_{\mathrm{T}}$ trigger particle, were suppressed when compared to those in \pp or \dAu collisions~\cite{Adams:2003im, Adler:2002tq}.
Comparison of the high $p_{\rm{T}}$ hadron yield in \AuAu collisions to the yield in binary-scaled \pp~collisions, provided evidence of suppression and energy loss of 
color-charged 
partons in the QGP due to increased medium-induced radiation and scattering within the medium along the 
parton shower.

The next generation measurements utilized reconstructed jets ({\textit i.e.}, groups of particles clustered via 
algorithms) that provided a better proxy for the initial hard-scattered parton's 
kinematics~\cite{Sterman:1977wj, Moretti:1998qx, Blazey:2000qt, Ellis:2007ib, Cacciari:2011ma, Salam:2009jx, 
Ali:2010tw}, and which facilitated direct comparisons to calculations and models that implemented partonic energy 
loss in the medium~\cite{Gyulassy:2003mc, Wang:2005sv, Wiedemann:2009sh, Zapp:2012ak}. Jet nuclear 
modification factors and coincidence measurements at RHIC~\cite{Adamczyk:2013jei, Adare:2015gla, 
Adamczyk:2016fqm, Timilsina:2016sjv, Adamczyk:2017yhe, Osborn:2018wxd, Adam:2020wen} and the 
LHC~\cite{Aad:2010bu, Chatrchyan:2011sx, Chatrchyan:2012nia, Aad:2012vca, Abelev:2013kqa, Aad:2013sla, 
Aad:2014wha, Aad:2014bxa, Adam:2015ewa, Adam:2015doa, Adam:2015mda, Khachatryan:2016jfl, Aaboud:2018twu, 
Aaboud:2019oop, Acharya:2019jyg} also showed suppression of the jet yield, implying jet energy loss. Since jets 
are collective objects, the third generation of measurements studied modifications to the jet structure via 
the jet fragmentation functions~\cite{Chatrchyan:2012gw, Adamczyk:2013jei, Chatrchyan:2014ava, Aad:2014wha, 
Aaboud:2017bzv, ATLAS:2017pgl, Sirunyan:2018qec, Aaboud:2018hpb, Aaboud:2019oac}, 
jet shapes~\cite{Chatrchyan:2013kwa, Khachatryan:2015lha, Khachatryan:2016tfj, Khachatryan:2016erx, 
Sirunyan:2018ncy}, jet mass~\cite{Acharya:2017goa, Sirunyan:2018gct, ATLAS-CONF-2018-014}, and also to their 
substructure via modified splittings~\cite{Sirunyan:2017bsd, Acharya:2019djg, ATLAS-CONF-2019-056}.
From di-hadron, jet-hadron and jet structure measurements, we observed that quenched jets have significantly 
enhanced (suppressed) yields of low (high) momentum constituents. The enhanced soft-constituents were found farther 
away from the jet axis, {\textit i.e.}, the periphery of the jets. This has been recently understood as an important 
signature of the medium response~\cite{Blaizot:2015lma, Qin:2015srf, Mehtar-Tani:2016nqm, Connors:2017ptx, 
Cao:2020wlm}.

The modifications to the jet structure indicate the essential need for studying jet quenching with a 
more differential approach. Since jet evolution in vacuum is characterized by a momentum fraction and 
opening angle as given by the DGLAP equations~\cite{Gribov:1972ri, Dokshitzer:1977sg, Altarelli:1977zs}, 
one might expect jet quenching mechanisms to depend on the jet shower topology~\cite{Mehtar-Tani:2016aco, 
Apolinario:2020uvt, Mehtar-Tani:2021fud}. Inclusive jet quenching measurements essentially integrate over 
all jet shower topologies. On the other hand, differential measurements of energy loss for jets, when tagged based on their shower 
topology via a substructure observable, can potentially isolate the varying mechanisms of jet quenching, 
which in turn leads to further stringent constraints on QGP transport parameters. 
In addition to energy loss being related to path length through the medium~\cite{Baier:1996kr, Baier:1996sk,  
Zakharov:1996fv, Zapp:2005kt, Zapp:2011ya, Burke:2013yra}, jet-medium interactions could also depend on 
the coherence length of the medium~\cite{CasalderreySolana:2012ef, Mehtar-Tani:2014yea}. Recent theoretical 
studies have shown that, depending on medium coherence length~\cite{Caucal:2021cfb, Caucal:2021lgf, Caucal:2021bae}, jets with similar kinematics but with a 
two-prong vs. one-prong structure (where a prong is a jet-like object within a jet) lose significantly 
different amounts of energy~\cite{Mehtar-Tani:2014yea, Mehtar-Tani:2016aco, Chien:2016led, Apolinario:2017qay, 
Milhano:2017nzm, Mehtar-Tani:2019ygg, Adhya:2019qse, Andres:2020vxs}. This dependence is a direct consequence 
of the transverse separation distance between the two prongs, at the time of 
formation, being greater or less than the 
resolution/coherence length of the medium~\cite{Mehtar-Tani:2014yea, Mehtar-Tani:2016aco}. Since the latter cannot be directly measured, the subjet angular 
separation is chosen as its proxy.\footnote{Subjet separation distance at formation is approximately 
$c ~ \tau_{f} ~ \theta$, where $\tau_{f}$ is the formation time and $\theta$ is the measured two-prong 
opening angle.} This resolving power is a direct application of the jet opening angle being greater or 
smaller than the coherence length. 

We calculate the di-jet asymmetry for trigger and recoil jets (see Fig.~\ref{fig:dijetselection}), 
defined as

\begin{equation}
\label{eq:ajequation}
A_{\rm{J}} \equiv \frac{p^{\rm{trigger}}_{\rm{T, jet}} - p^{\rm{recoil}}_{\rm{T, jet}}}{p^{\rm{trigger}}_{\rm{T, jet}} + p^{\rm{recoil}}_{\rm{T, jet}}},
\end{equation}
as a quantitative measure of jet quenching effects for jets of varying angular scales.
For di-jet selections in \AuAu collisions at RHIC, the highest $p_{\rm{T}}$ jet is biased to originate 
preferentially from production vertices in the periphery of the overlap region of the colliding 
ions~\cite{Renk:2012hz}, often referred to as a surface bias.
Since the quenched energy of these recoil jets is found to be contained in soft particles 
(with $0.2 < p_{\rm{T}} < 2$ GeV/$c$) that are distributed around the jet axis~\cite{Adamczyk:2013jei}, 
the dependence of quenching on the angular scales can be studied by comparing the same measurement for 
narrow- vs.~wide-angle jets. If the jet opening angle, as defined by a substructure observable, 
is smaller than 
the medium coherence length, the two prongs (hereafter referred to as subjets) are unresolved by the 
medium and will be quenched as an effective single color-charge~\cite{Mehtar-Tani:2014yea}. 
On the other hand, if the jet opening angle is larger than the coherence length, the two prong remain resolved within the medium  and as a result, two color charges undergo independent 
interactions with the medium, resulting in larger energy loss of their parent parton. 
These differential types of measurements constitute a first attempt at tagging jet populations 
impacted by varying energy loss mechanisms such as medium induced radiation, as in the coherent/de-coherent 
picture, or medium induced splittings that result in modifications to the jet substructure.    

The paper is organized as follows. Sect.~\ref{sec:dataset} introduces the STAR detector and the datasets used 
in this analysis along with the jet reconstruction parameters. The measurements 
of SoftDrop observables are presented in Sect.~\ref{sec:SoftDrop} including a study of the sensitivity of 
these observables to the 
underlying event in heavy-ion collisions. In Sect.~\ref{sec:twosubjet} measurements of fully 
corrected two-subjet 
observables in \pp collisions are introduced and compared to leading order Monte Carlo (MC) simulations. 
Their observability in the heavy-ion collision environment is also discussed. 
The measurements of the subjet observables in \AuAu collisions are presented in
Sect.~\ref{sec:diffAj} along with differential measurements of the di-jet asymmetry for recoil 
jets with varying opening angle. A summary and discussion of the implications of these findings and 
an outlook to the future regarding such differential measurements
are given in Sect.~\ref{sec:summary}.

\section{Analysis details}
\label{sec:dataset}

\subsection{STAR Detector}

STAR is a large, multipurpose detector at RHIC~\cite{Ackermann:2002ad}. It consists of a solenoidal magnet 
and multiple detectors used for triggering, tracking, particle identification and calorimetry. This analysis 
uses charged tracks from the Time Projection Chamber (TPC)~\cite{Anderson:2003ur} and neutral energy recorded 
by the Barrel Electromagnetic Calorimeter (BEMC)~\cite{Beddo:2002zx}. Charged tracks assigned to the primary 
vertex are required to have a global distance of closest approach (DCA) to the primary vertex of less than 
1 cm, greater or equal to $20$ fit points in the TPC, and at least 52\% of the maximum number of fit points possible to 
eliminate split tracks. The BEMC consists of 4800 individual towers, with a tower size of $0.05\times0.05$ in $\eta \times \phi$. Any event with a $p_{\rm{T}} > 30$ GeV/$c$ charged-particle track is discarded due to poor 
momentum resolution, and to be consistent, events with BEMC tower $E_{\rm{T}} > 30$ GeV are also rejected. 
To avoid the double-counting of charged-particle track energy and the energy deposition in the corresponding 
matched BEMC tower, the full track energy (assuming pion mass) is subtracted from the tower energy. This 
procedure is referred to as a 100\% hadronic correction. This approach has been used in past STAR 
publications~\cite{Adamczyk:2016fqm, Adam:2019aml, Adam:2020kug} with the desirable effect of providing 
good jet energy resolution. Any towers with $E_{\rm{T}} < 0$ GeV after hadronic correction are ignored in 
the analysis.

\subsection{Dataset and jet reconstruction}

We analyze \AuAu collisions and the corresponding \pp reference data at $\sqrt{s_{\rm{NN}}} = $ 200 GeV 
recorded by STAR in 2007 and 2006, respectively, with a high tower (HT) trigger requiring at least one online calorimeter 
tower with $E_{\rm{T}} > 5.4$ GeV. The triggered data corresponds to an integrated luminosity of 11.3 pb$^{-1}$ and 
0.6 nb$^{-1}$ for \pp and \AuAu collisions, respectively. We also utilize \pp data at the same center of mass energy 
taken in 2012 \footnote{This dataset and its corresponding simulations and detector corrections are identical in procedure to a previous STAR publication~\cite{Adam:2020kug}} for the fully corrected measurements. Events from all datasets are required to have 
the primary vertex within 30 cm of the center of the TPC detector along the beam direction. Centrality 
 in \AuAu events estimates the overlap of the two colliding nuclei and is determined 
by the raw charged-particle multiplicity in $|\eta| < 0.5$ within the TPC~\cite{Adamczyk:2016fqm}.
We report results for the most central (0-20\%) \AuAu collisions.

\begin{SCfigure}[][h] %  figure placement: here, top, bottom, or page
   \centering
   \includegraphics[width=0.5\textwidth]{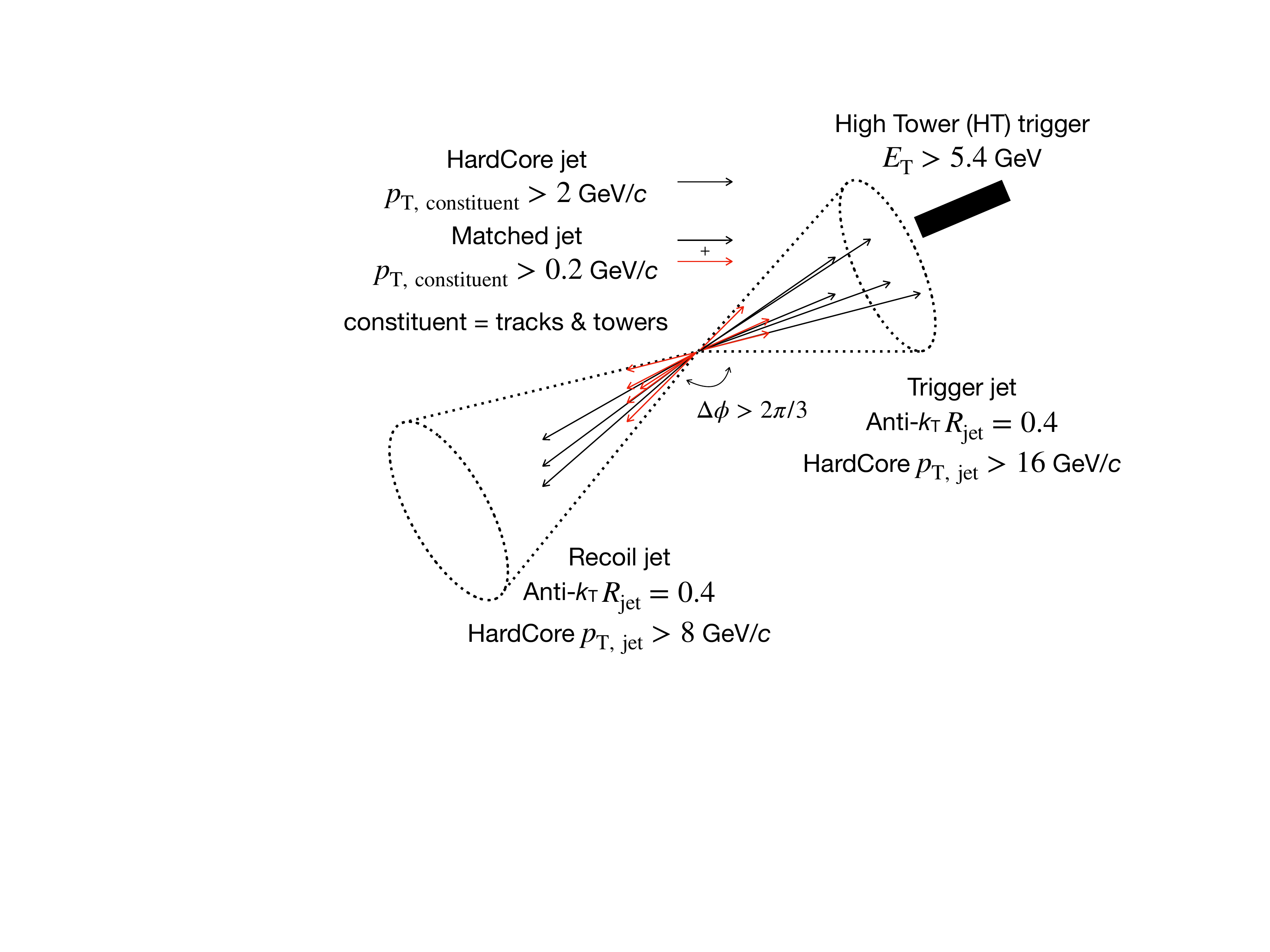}
   \caption{Description of the di-jet event selection noting the trigger and recoil jets along with the HardCore and Matched jets which include the hard and soft constituents in the black and red colored arrows, respectively. The high tower trigger is shown in the thick black shaded region in the trigger jet region. The thrust axes for HardCore and Matched jets can be slightly different since they are associated via $\Delta R$ matching criterion.}
   \label{fig:dijetselection}
\end{SCfigure}

In HT triggered \AuAu collisions we utilize two separate jet collections, both clustered on tracks and towers (denoted as Ch$+$Ne in the figures) with the 
anti-$k_{\rm{T}}$ algorithm~\cite{Cacciari:2008gp, Cacciari:2011ma} with the resolution parameter $R = 0.4$, called HardCore and Matched di-jets. HardCore 
jets are clustered with constituents (tracks and towers) with $2 < p_{\rm{T}} ~ 
(E_{\rm{T}}) < 30$ GeV/$c$ (GeV), and Matched jets utilize constituents with 
$0.2 < p_{\rm{T}}(E_{\rm{T}}) < 30$ GeV/$c$ (GeV) and are geometrically matched to the HardCore 
jets~\cite{Adamczyk:2016fqm} as shown in Fig.~\ref{fig:dijetselection}. Due to the high $p_{\rm{T}}$ constituent threshold, the HardCore jets are effectively free of combinatorial background.  
The background is subtracted 
from the Matched jets via the constituent subtraction method~\cite{Berta:2014eza}, with parameters 
$\alpha = 2$ and the maximum allowed distance between a particle and a ghost is set to $1.0$ in order 
to suppress the underlying event contribution to the jet. Di-jet pairs are selected with the 
trigger HardCore jet ($\Delta R_{\mathrm{jet,HT}} = \sqrt{(\phi_{\mathrm{jet}} - \phi_{\mathrm{HT}})^{2} 
+ (\eta_{\mathrm{jet}} - \eta_{\mathrm{HT}})^{2}} < 0.4$) and recoil HardCore jet 
($\Delta \phi_{\rm{jet, HT}} > 2\pi/3$), both having minimum jet 
$p_{\rm{T}}$ requirements ($p^{\rm{trigger}}_{\rm{T, jet}} > 16$ GeV/$c$, 
$p^{\rm{recoil}}_{\rm{T, jet}} > 8$ GeV/$c$). 
We also require that the trigger HardCore jet has larger transverse momenta than 
the recoil HardCore jet which results in the di-jet asymmetry as defined in Eq.~\ref{eq:ajequation} 
to be positive for HardCore di-jets. We do not impose any momentum thresholds on the Matched di-jets 
and as a result the Matched $A_{\rm{J}}$ can be positive or negative. 

To make meaningful comparisons to the \AuAu data, the \pp reference events are embedded into the 0-20\% most-central 
\AuAu data from the minimum-bias dataset \footnote{The STAR minimum-bias trigger requires an event to have signals in both forward scintillator Vertex Position Detectors (VPD), along with a requirement of at least one neutron in each Zero Degree Calorimeter (ZDC). This biases towards a requirement of hadronic interaction of both the Au-ions.} to capture the effect of the large fluctuations in background energy density in high
 multiplicity \AuAu collisions. The embedded reference is denoted as \ppAA henceforth. The relative difference 
in detector performance is taken into account since the TPC experiences a degradation in efficiency as the 
event multiplicity increases. Thus, charged-particle tracks in a \pp event are artificially removed according 
to the tracking efficiency ratio between \AuAu and \pp collisions, which is approximately $90\%$ for most track momenta 
and rapidity. This procedure consequently enables a direct comparison of the \pp data to \AuAu data at the 
detector-level, including the effects of the fluctuating underlying event. A description of the detector 
effects necessary to compare a model calculation with the measured data at the detector-level is provided 
in the Appendix. 

Systematic uncertainties in the embedded \ppAA reference are indicated by the shaded region and are estimated by the difference in detector performance for both the TPC and the BEMC between the two datasets as was done in the previous publication~\cite{Adamczyk:2016fqm}. The ratio of the TPC tracking efficiencies for \AuAu and \pp collisions is approximately $90 \pm 7\%$. The effect of this systematic uncertainty is assessed by repeating the \pp embedding procedure with the respective minimum ($83\pm$) and maximum ($97\pm$) relative efficiency. The uncertainty due to the tower gain is assessed by repeating the measurement and varying the energy of all towers by $\pm 2$\%. The resulting uncertainty from the tower energy scale is negligible compared to that of the tracking efficiency. The final reported uncertainties are the quadrature sum of the relative tracking efficiency uncertainty and the relative tower energy scale uncertainty. There are no systematic uncertainties on the \AuAu data presented in this measurement since it is at the detector-level and thus uncertainties are only presented for our embedded reference.  

\section{SoftDrop jet substructure}
\label{sec:SoftDrop}

\begin{SCfigure}[][h] %  figure placement: here, top, bottom, or page
   \centering
   \includegraphics[width=0.5\textwidth]{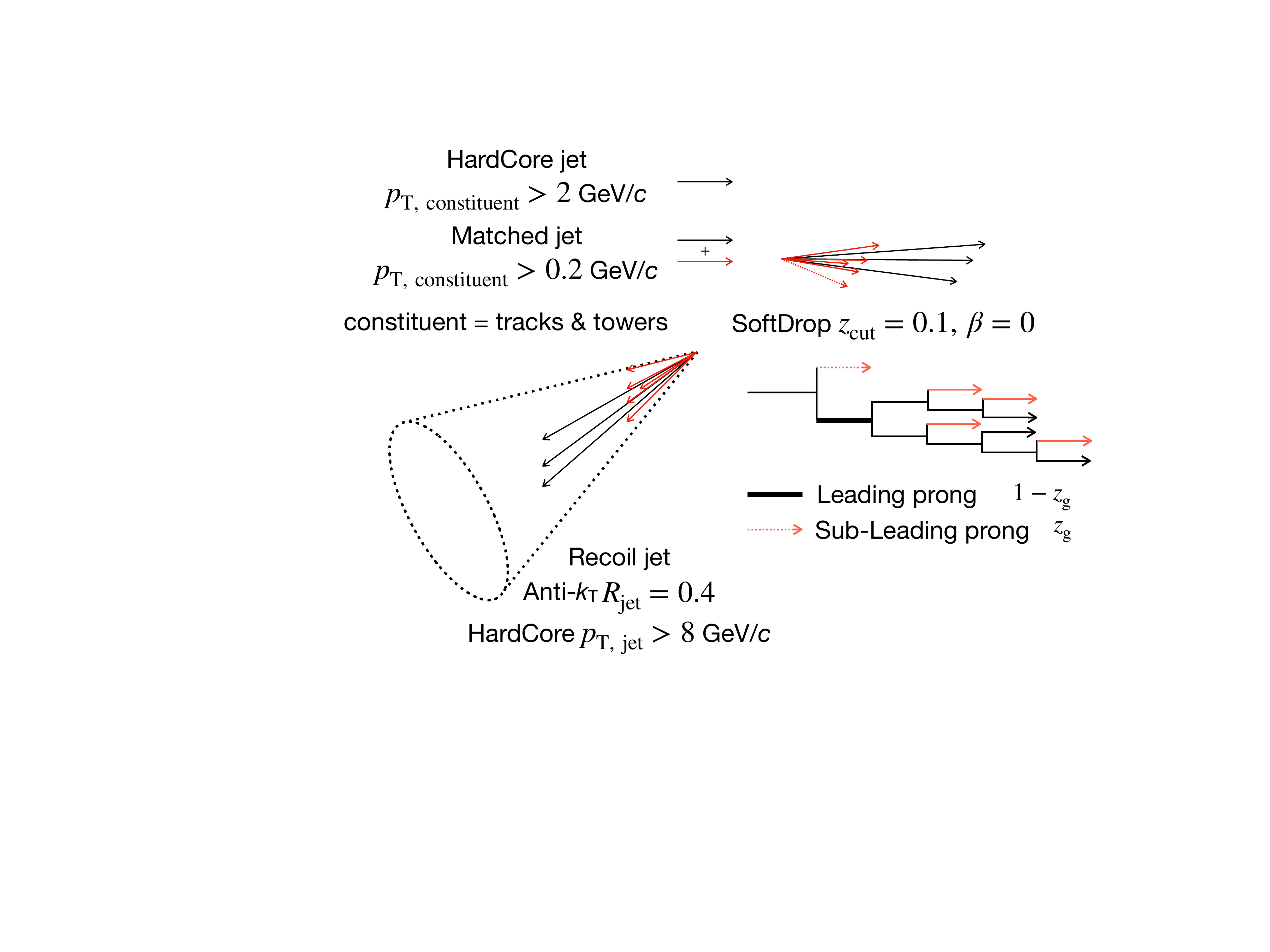}
   \caption{Visualization of the SoftDrop de-clustering step applied on a recoil jet resulting in the selection of the leading and subleading prongs.}
   \label{fig:sdcartoon}
\end{SCfigure}

The SoftDrop~\cite{Dasgupta:2013ihk,Larkoski:2014wba} grooming algorithm removes soft, wide-angle radiation from a sequentially clustered jet. This is achieved by recursively de-clustering the jet's branching history and discarding prongs until the transverse momenta $p_{\rm{T,1}}, p_{\rm{T,2}}$ of the current pair of prongs fulfill the \mbox{SoftDrop} conditions,

\begin{equation}
\label{eq:sdequation}
\begin{split}
z_{\rm{g}} & = \frac{\min(p_{\rm{T,1}},p_{\rm{T,2}}) }{ p_{\rm{T,1}}+p_{\rm{T,2}}} > z_{\text{cut}}\left( \frac{R_{\rm{g}}}{R_{\rm{jet}}}\right)^\beta, \\
R_{\rm{g}} & = \Delta R_{\rm{1, 2}} = \sqrt{\Delta \eta^{2}_{\rm{1, 2}} + \Delta \phi^{2}_{\rm{1, 2}}},
\end{split}
\end{equation}
where $R_{\rm{g}}$ is the opening angle between the two prongs and $R_{\rm{jet}}$ is the jet resolution parameter. The current analysis sets $\beta=0$, and we adopt the default choice $z_{\text{cut}}=0.1$~\cite{Larkoski:2015lea}. In \pp collisions, the SoftDrop grooming procedure results in reducing the non-perturbative contributions to the jet which aids in translating the splitting in a jet tree to a partonic splitting via the DGLAP splitting functions. The two SoftDrop observables, \zg~and \rg, were shown to be sensitive to the momentum and angular scales in \pp collisions in a recent STAR publication~\cite{Adam:2020kug}, wherein the data were compared to both MC event generators and theoretical calculations. Measuring the \zg~in heavy-ion collisions opens up the possibility of studying modifications to the splitting kernel, a characteristic aspect in some energy loss models~\cite{CasalderreySolana:2012ef, Mehtar-Tani:2014yea}, but it could also indicate changes due to quenching of the subjets after a vacuum-like fragmentation. A diagrammatic representation of the SoftDrop algorithm on a recoil jet is shown in Fig.~\ref{fig:sdcartoon}, which highlights the de-clustered tree and the leading and subleading prongs which are utilized in the measurements of \zg~and \rg. It is important to note that the first selected split could indeed be different for HardCore and Matched jets due to the inclusion of the softer components of the jet.  

\begin{figure}[h] %  figure placement: here, top, bottom, or page
   \centering
   \includegraphics[trim=120 75 0 20,clip,width=0.75\textwidth]{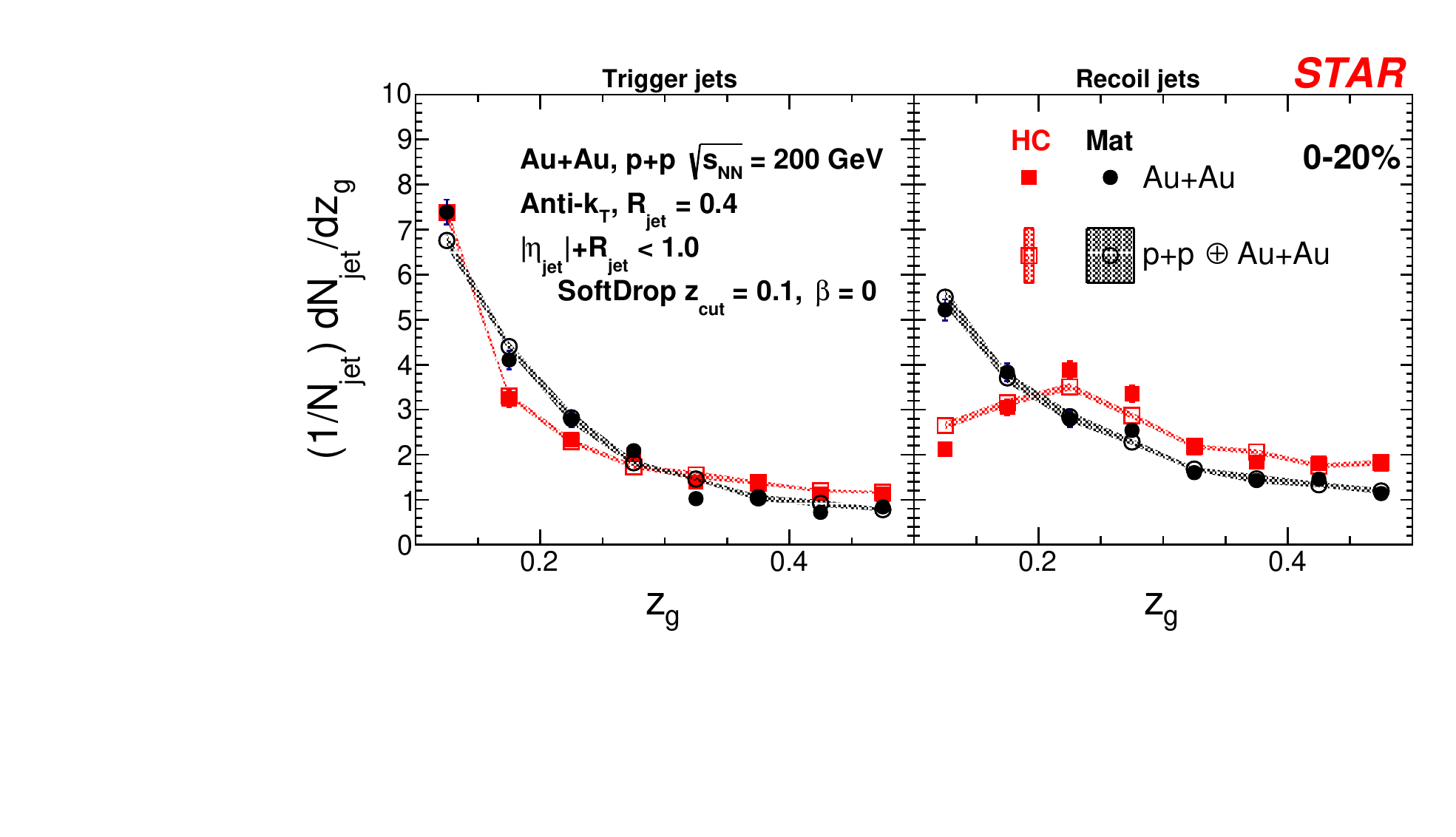}
   \includegraphics[trim=120 75 0 20,clip,width=0.75\textwidth]{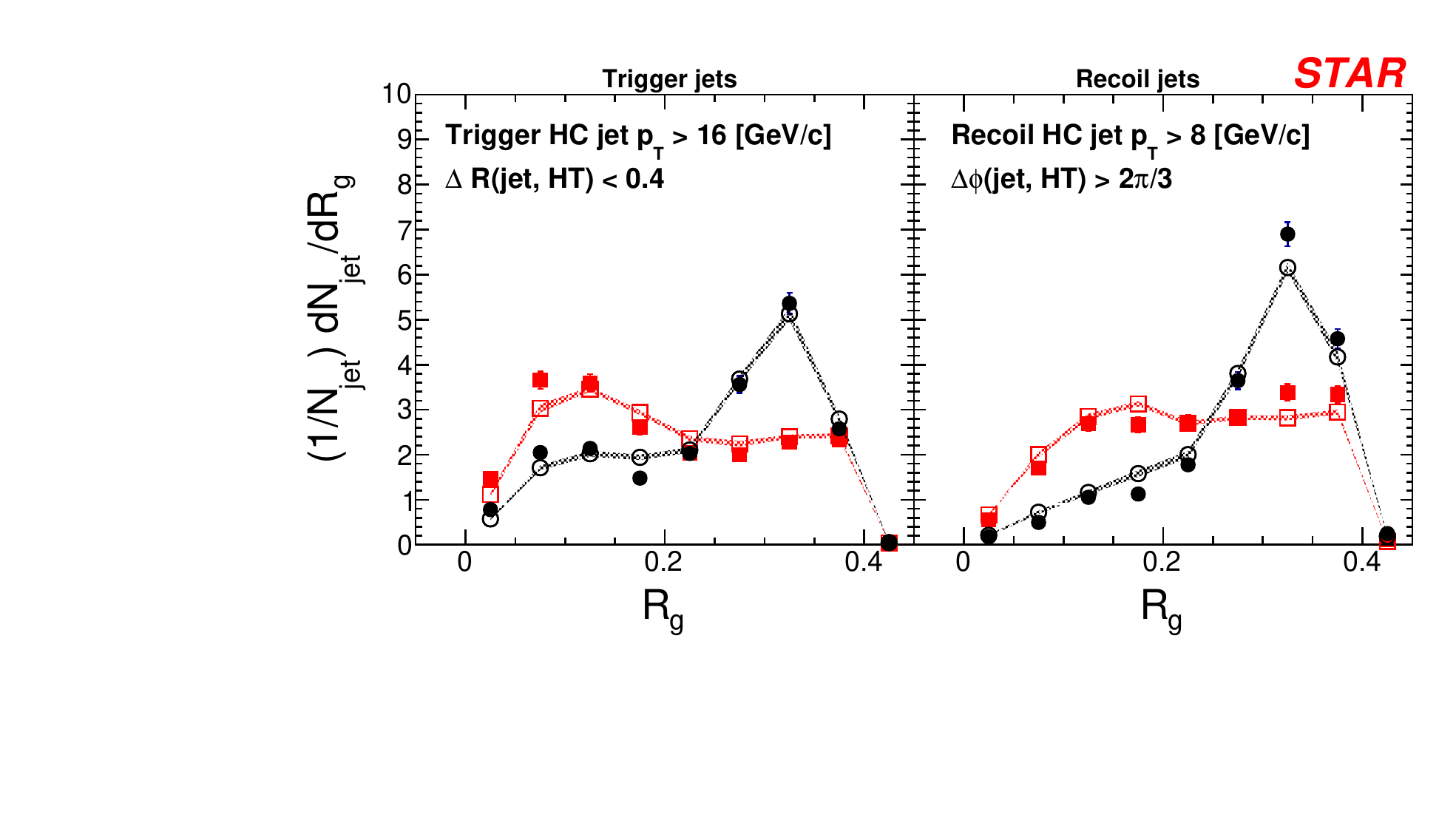}
   \caption{Distributions of softDrop \zg~(top) and \rg~(bottom) for trigger (left) and recoil (right) jets for \AuAu data (filled symbols) and the  \ppAA reference (open symbols). The shaded regions represent systematic uncertainty in the embedded reference. The red and black points represent HardCore and Matched jets, respectively.}
   \label{fig:zgrg}
\end{figure}

Figure~\ref{fig:zgrg} shows the SoftDrop \zg~(top) and \rg~(bottom) for trigger (left) and recoil (right) jets with $p_{\rm{T}} > 16$ GeV/$c$ and  $p_{\rm{T}} > 8$ GeV/$c$, respectively.  The \AuAu data are represented by solid symbols compared to the \ppAA results in open markers. HardCore (Matched) jets are shown in the red squares (black circles). For both \zg~and \rg~we find no significant difference in the shape of the distributions between \AuAu and \ppAA as a consequence of jet quenching. The apparent peak in the HardCore \zg~distribution (as seen in the red markers in the top right panel) is due to a kinematic constraint arising from a jet constituent threshold of $2$ GeV/$c$. We observe significant differences in the \rg~distribution between HardCore and Matched jets (both trigger and recoil) where the latter peaks at values closer to $0.33$. Given that HardCore jets are less affected by the combinatorial background, the Matched jet \rg~shows its sensitivity to the underlying event. The $0.33$ peak of the Matched jet \rg~is indicative of a geometric selection of particles at the edge of an $R=0.4$ jet that pass the grooming requirement and are contained within the jet.

\begin{figure}[h] %  figure placement: here, top, bottom, or page
   \centering
   \includegraphics[width=0.8\textwidth]{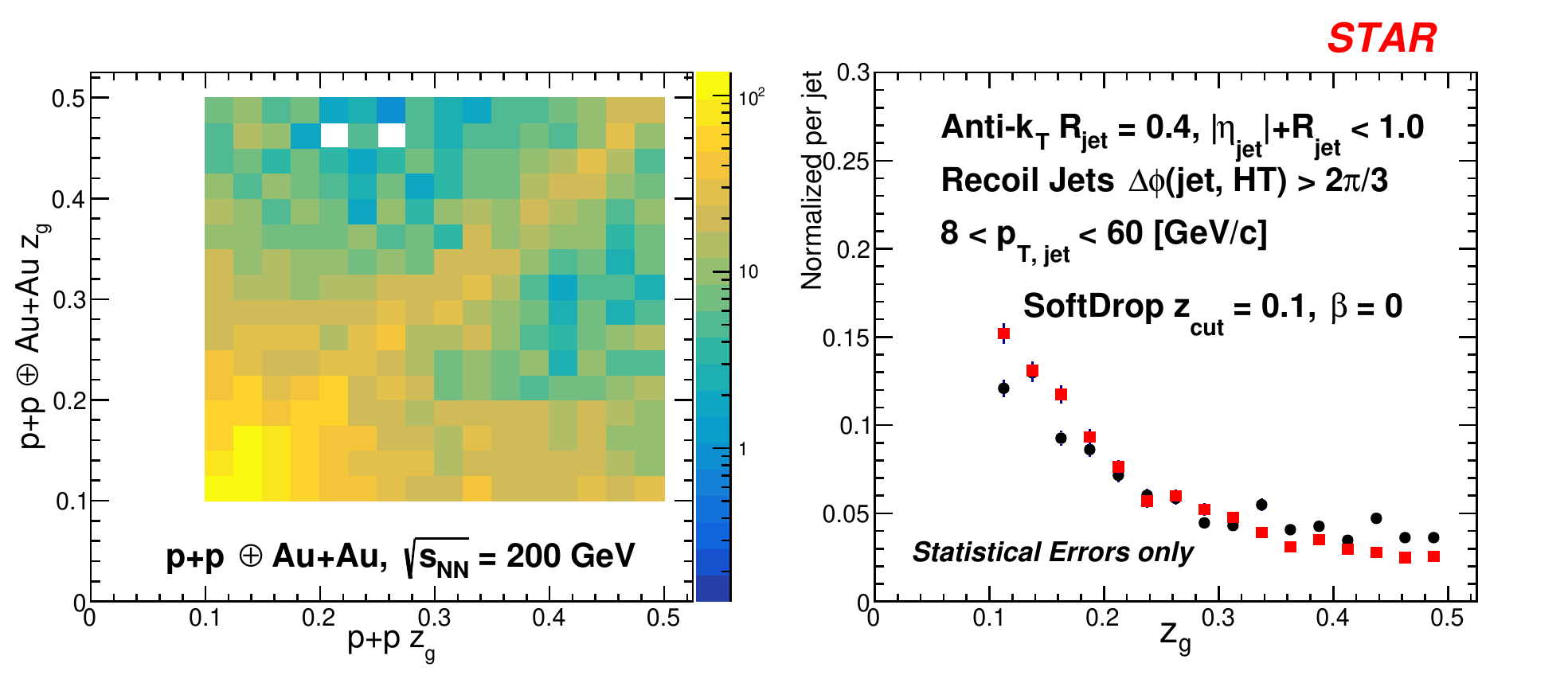}
   \includegraphics[width=0.8\textwidth]{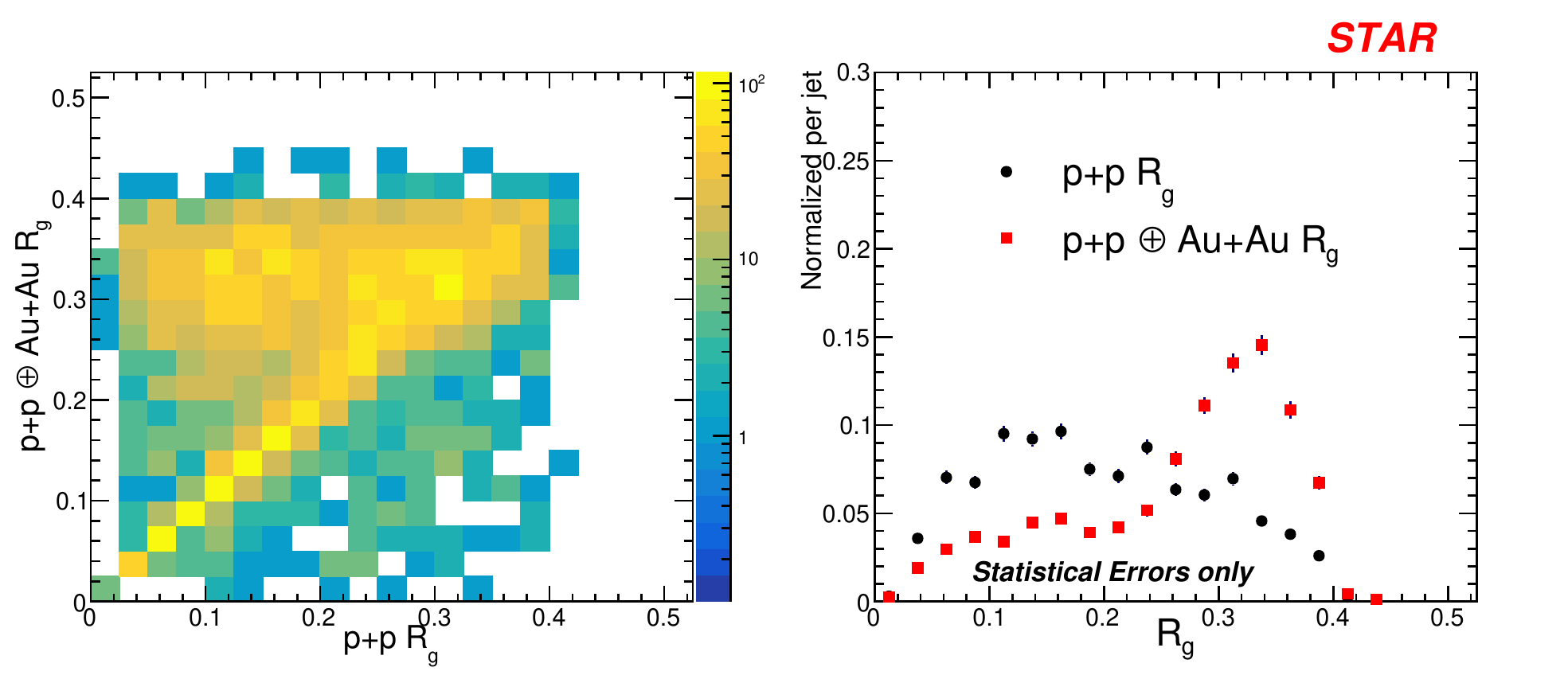}
   \caption{Left panels: Correlation studies due to the heavy-ion underlying event for SoftDrop \zg~(top) and \rg~(bottom). Right panels: The projections along x- and y-axis shown with only statistical errors.}
   \label{fig:zgrgEmbeffect}
\end{figure}

We further quantify the effect of the underlying event on SoftDrop observables by embedding \pp events into minimum-bias \AuAu events. Jets are found in both \pp and \ppAA embedding data and these jets are then matched geometrically. The SoftDrop observables in the two datasets are compared via a 2D correlation. The left panels in Fig.~\ref{fig:zgrgEmbeffect} show this correlation where the $x$-axis represents the observable from the \pp jet and the y-axis the corresponding value for the embedded, constituent subtracted \ppAA jet. In the right panels, individual projections of the \zg~(top right) and \rg~(bottom right) are shown with the \pp data in filled black circles and the \ppAA data in filled red boxes.  

While the 1-dimensional \zg~distributions in the two datasets are similar, we do not observe a significant diagonal correlation between substructure observables. The effect of the underlying event is prominent in the \rg, where the correlation shows a particular failure mode of the SoftDrop algorithm (with default parameters $z_{\rm{cut}} = 0.1$ and $\beta = 0$) in which particles from the underlying event, that are not correlated with the jet by definition, make it through the grooming procedure and end up being selected as the first, hardest split. One can indeed modify the grooming procedure to reduce the non-diagonal component in the correlation by varying the grooming criterion~\cite{Mulligan:2020tim}. Since we aim to tag jets based on their inherent angular scales via an opening angle, the subsequent bias on the surviving jet population due to a momentum fraction threshold essentially gets varied and convoluted for jets of varying jet kinematics. Thus, we present a new class of observables that characterizes jet substructure via reclustered subjets.  

\section{Momentum and angular scales via subjets}
\label{sec:twosubjet}

\begin{SCfigure}[][h] %  figure placement: here, top, bottom, or page
   \centering
   \includegraphics[width=0.5\textwidth]{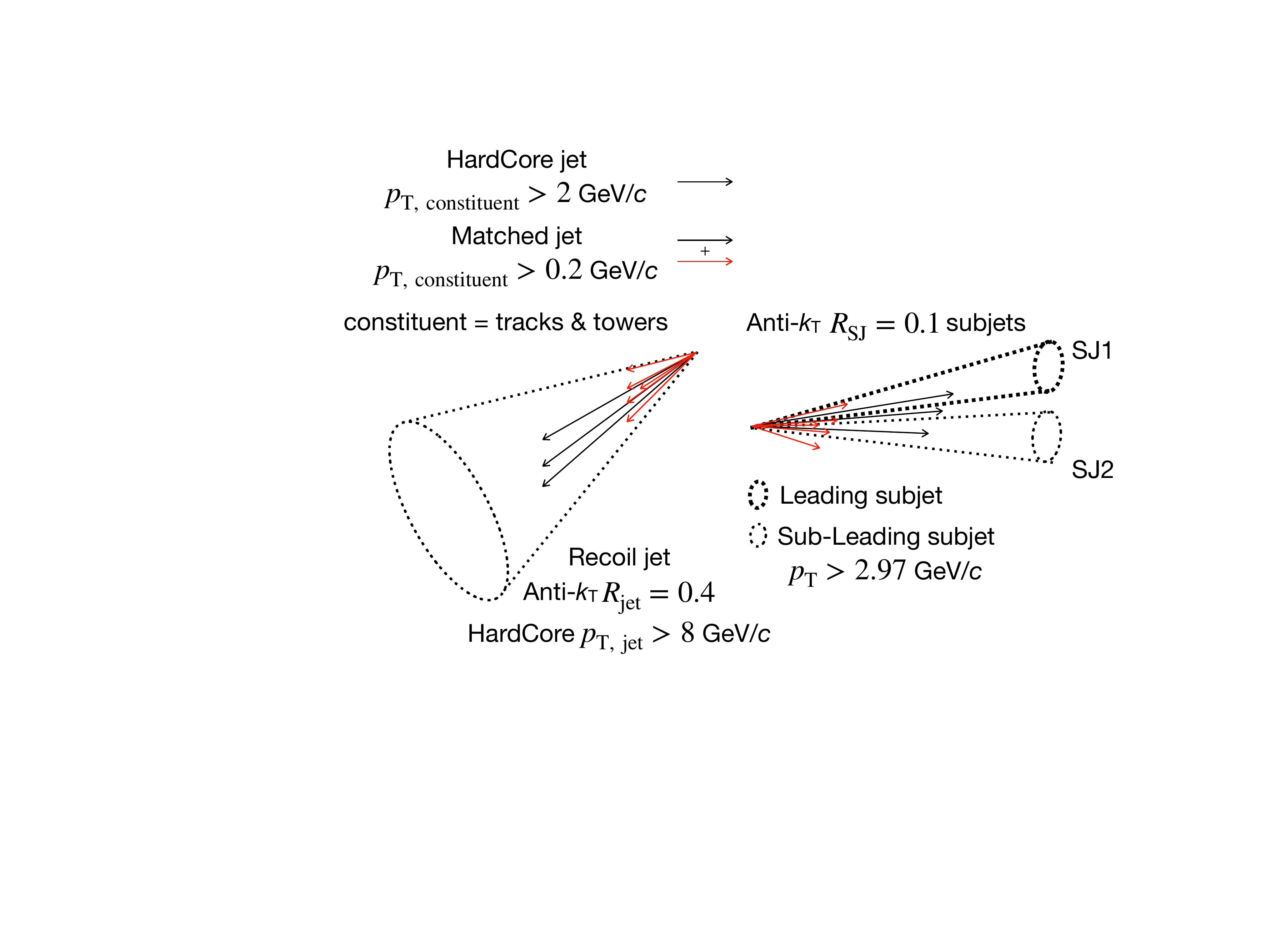}
   \caption{Visualization of the subjet definition and the selection of the leading and subleading subjets which are utilized in the selection of narrow/wide jets. The thrust axes for HardCore and Matched jets can be slightly different since they are associated via $\Delta R$ matching criterion.}
   \label{fig:sjcartoon}
\end{SCfigure}

\begin{figure}[h] %  figure placement: here, top, bottom, or page
   \centering
   \includegraphics[width=0.99\textwidth]{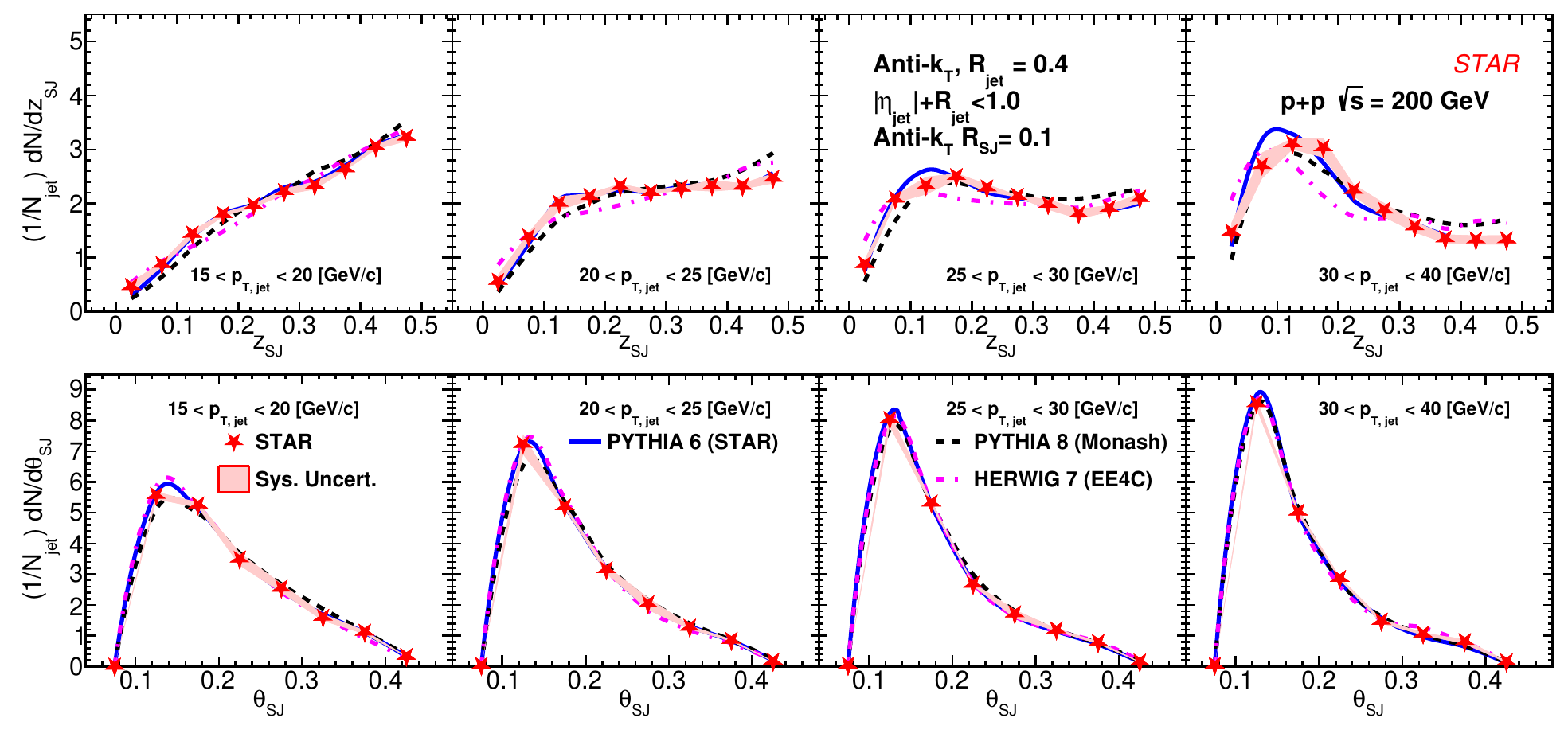}
   \caption{Distributions of fully unfolded two-subjet observables \zsj~in the top panels and \tsj~in the bottom panels for inclusive jets in \pp collisions shown in the red star markers, compared to leading order MC event generators PYTHIA 8 (black), PYTHIA 6 (blue) and Herwig 7 (magenta) for low-\pt jets on the left to high-\pt jets on the right. The shaded red regions represent the systematic uncertainties in the data points.}
   \label{fig:pptwosubjet}
\end{figure}

The idea of utilizing subjets to probe jet substructure has been explored recently in theoretical studies where subjets are considered as advantageous probes to quantify jet-medium interactions~\cite{Apolinario:2017qay}. We introduce and measure subjet observables for the first time where anti-$k_{\rm{T}} ~ R = 0.4$ jet constituents are reclustered with the anti-$k_{\rm{T}}$ clustering algorithm with smaller $R_{\rm{SJ}}$ and the two highest \pt subjets are considered as shown in Fig.~\ref{fig:sjcartoon}. The momentum fractions and opening angle are defined as follows,

\begin{equation}
\label{eq:sjequation}
\begin{split}
z_{\rm{SJ}} & = \frac{\min(p_{\rm{T,SJ1}}, p_{\rm{T, SJ2}})}{p_{\rm{T, SJ1}} + p_{\rm{T, SJ2}}}, \\
\theta_{\rm{SJ}} & = \Delta R (\rm{SJ1, SJ2}),
\end{split}
\end{equation}
where $\rm{SJ1}$ and $\rm{SJ2}$ are the leading and sub-leading subjets, respectively. The fully corrected subjet distributions, \zsj~(top) and \tsj~(bottom) in \pp\ collisions are measured for inclusive anti-$k_{\rm{T}}, ~ R=0.4$ jets. Figure.~\ref{fig:pptwosubjet} shows \zsj~and \tsj~with a subjet radius of $R_{\rm{SJ}}=0.1$ for various jet \pt selections increasing from left to right. The data in the red star markers (along with the total systematic uncertainty in the shaded red regions) were corrected via an iterative Bayesian unfolding technique as implemented in the RooUnfold framework~\cite{Adye:2011gm}. The correction procedures and systematic uncertainties are identical to the procedure outlined in a recent STAR publication~\cite{Adam:2020kug}.

There are four major sources of uncertainties, described in order of importance as the source and its corresponding variation in parentheses: tracking efficiency in \pp collisions ($\pm 4\%$), BEMC tower energy scale ($\pm 3.8\%$), hadronic correction ($50\% - 100\%$), and the unfolding correction. The systematic uncertainties due to the unfolding procedure include changing the iteration parameter $(2-6)$ and varying the shape of the prior distribution. The default prior in the unfolding procedure is PYTHIA 6~\cite{Sjostrand:2006za} with the STAR tune~\cite{Adam:2019aml, Adam:2020kug}, where PYTHIA 8~\cite{Sjostrand:2014zea} using the Monash tune and HERWIG 7~\cite{Bahr:2008pv, Bellm:2015jjp} using the EE4C tune serve as systematic variations.
The data are compared to leading order MC predictions such as PYTHIA 6 (blue), PYTHIA 8 (black) and HERWIG 7 (pink). The \zsj~distributions show small but significant differences for $p_{\rm{T, jet}} > 25$ GeV/$c$ with the MC predictions, whereas the \tsj~exhibit very good agreement between data and predictions. We also find a characteristic change in the shape of the \zsj~distribution as the jet \pt  increases. The mean value of the \tsj~shifts to smaller values and one of the subjets effectively captures the jet core. We also note that the \tsj~distribution is quantitatively similar for the different MC, which have different parton showers and hadronization models and accurately reproduce the distribution in data, which points to the stability of the \tsj~observable.

Before measuring these substructure observables in \AuAu collisions, we studied the impact of the underlying event by embedding PYTHIA 8 \pp events into \AuAu minimum-bias events. For a given subjet radius ($R_{\rm{SJ}} = 0.1$) we estimated an effective subjet \pt threshold in order to reduce sensitivity to the fluctuating background. A subjet \pt threshold of $p_{\rm{T}} > 2.97$ GeV/$c$ was calculated by taking the mean $(\mu)$ plus $3\sigma$ limit estimated from random cone studies\footnote{This procedure involves dropping circles of radius $R=0.1$ inside the PYTHIA 8 jets embedded in minimum-bias \AuAu events with a random center $\eta-\phi$ within the jet cone.}. It is important to note that the subjet selection criteria is based on absolute \pt as opposed to SoftDrop, which employs a momentum fraction ($z_{\rm{cut}}$) threshold. Given that these jets undergo interactions with the medium at a characteristic energy scale, a momentum fraction cut could impose a hitherto unknown, varying survivor bias in the selected jet population based on the jet kinematics. By selecting jets based on a subjet $p_{\rm{T, SJ}} > 2.97$ GeV/$c$ cut, the selection scale is constant with respect to jet momenta and the bias in the surviving jet population can be theoretically quantified. 

\begin{figure}[h] %  figure placement: here, top, bottom, or page
   \centering
   \includegraphics[width=0.8\textwidth]{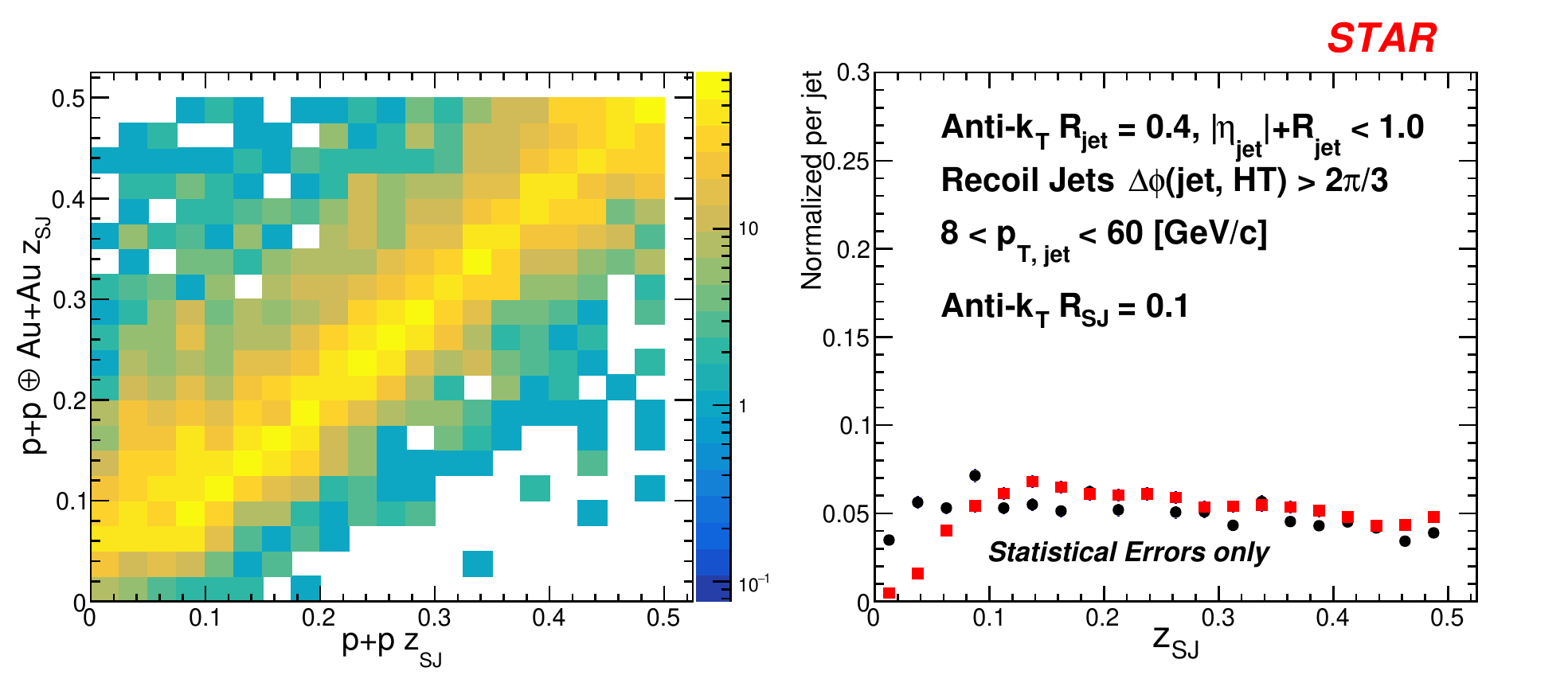}
   \includegraphics[width=0.8\textwidth]{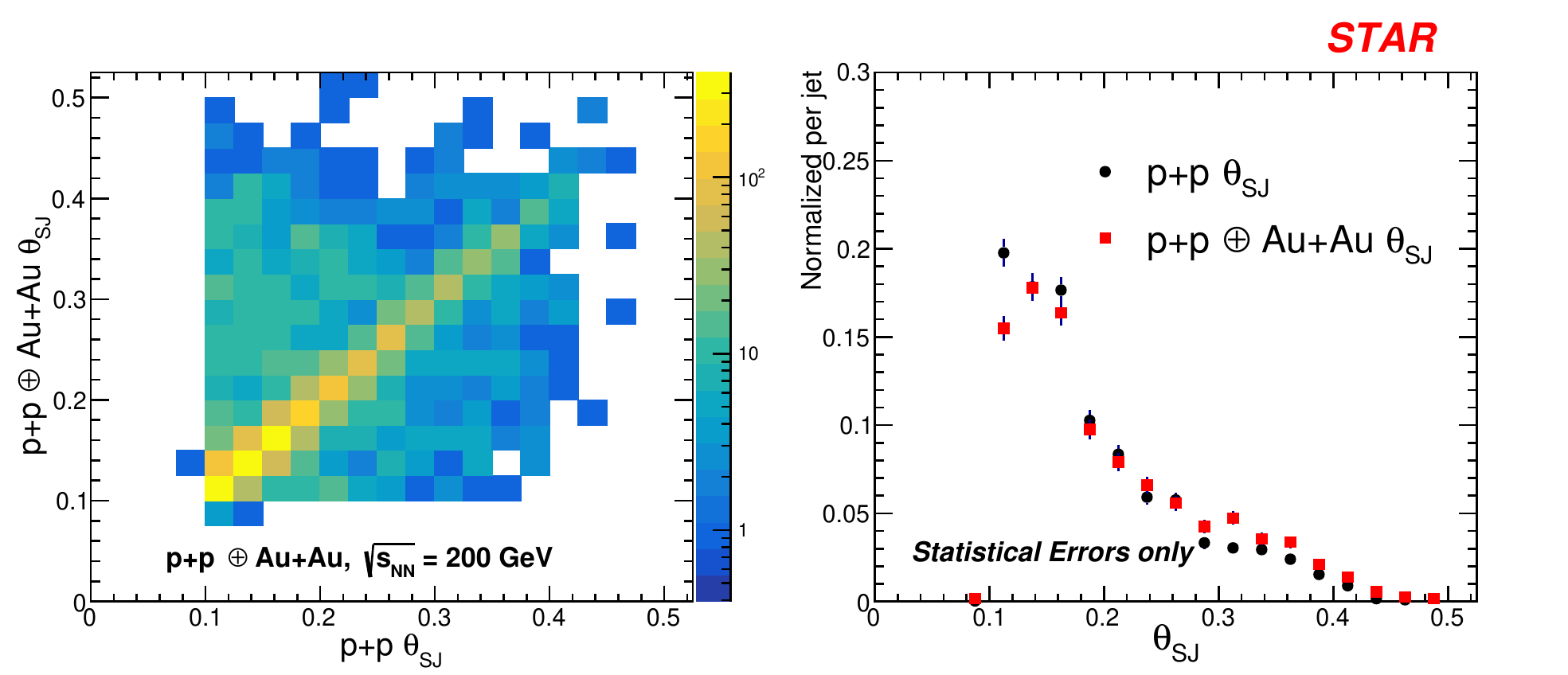}
   \caption{Left panels: Correlation studies of the two subjet observables \zsj~(top row) and \tsj~(bottom row) similar to Figure~\ref{fig:zgrgEmbeffect}. Right panels:  The projections along x- and y-axis shown with only statistical errors.}
   \label{fig:zsjtsjEmbeffect}
\end{figure}

The effects of the underlying event on the two subjet observables are also explored in a procedure identical to the one described in the previous section. Figure~\ref{fig:zsjtsjEmbeffect} shows the correlation matrix on the left and the individual distributions on the right for \zsj~(top row) and \tsj~(bottom row). In contrast to SoftDrop observables, we find that both subjet observables, \zsj~and \tsj, have a relatively strong diagonal correlation along with an overlap in the distributions themselves as shown in the right panels of Fig.~\ref{fig:zsjtsjEmbeffect}. The robustness of the two subjet observables facilitates a direct comparison between \AuAu and \ppAA results where the effect of the underlying event is effectively negated.  

\begin{figure}[h] %  figure placement: here, top, bottom, or page
   \centering
   \includegraphics[trim=120 75 0 20,clip,width=0.75\textwidth]{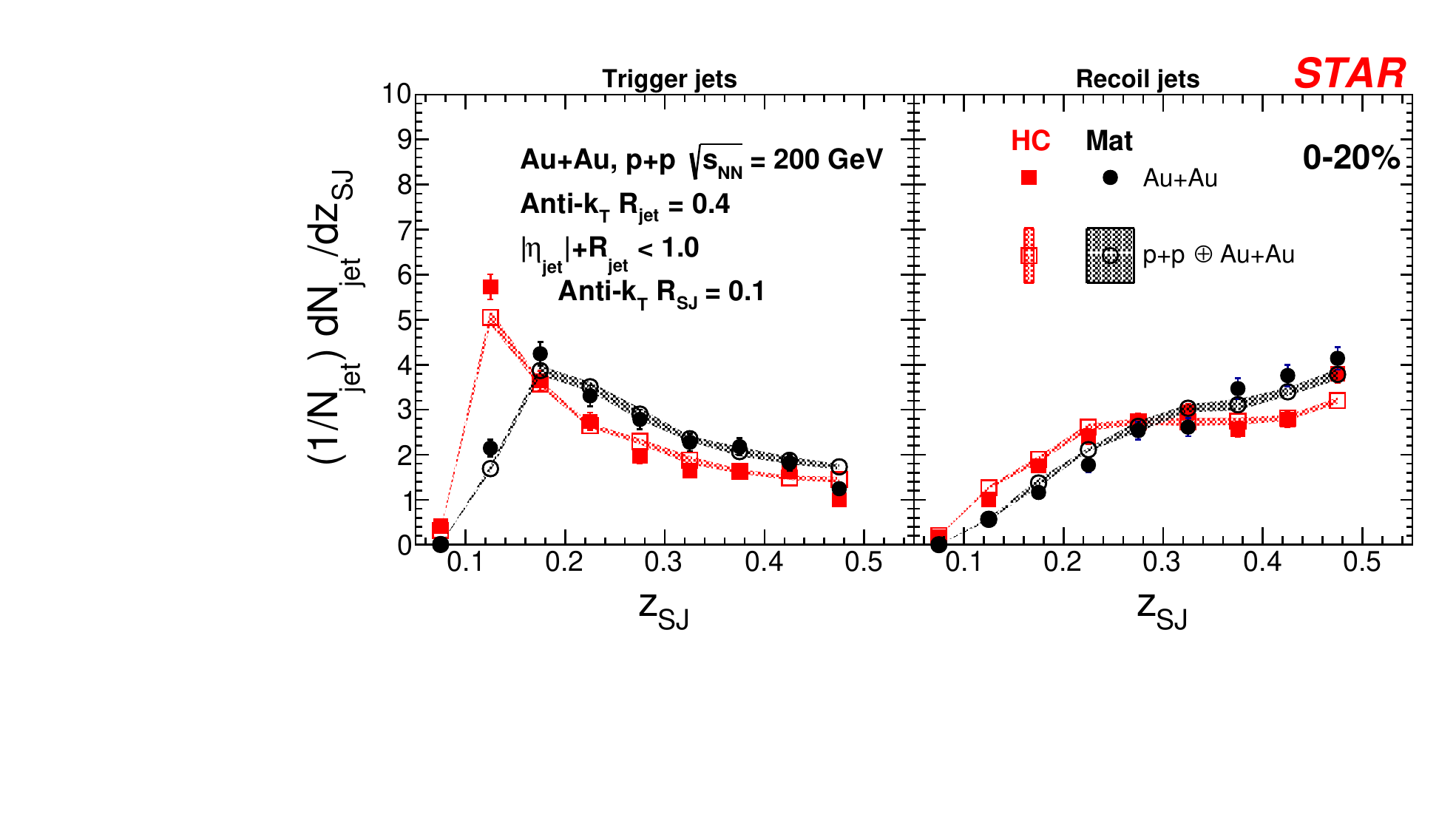}
   \includegraphics[trim=120 75 0 20,clip,width=0.75\textwidth]{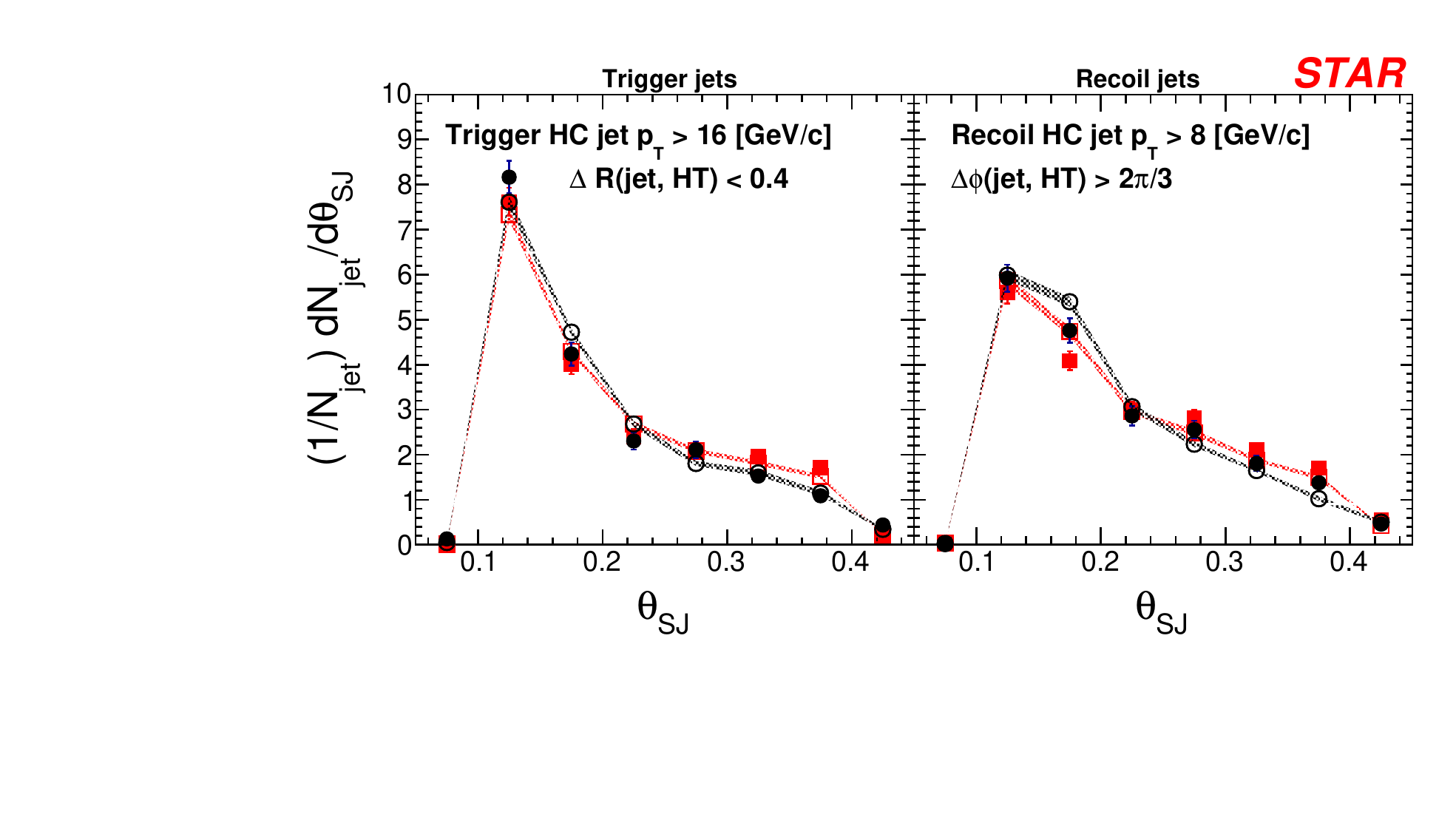}
   \caption{Distributions of subjet observables \zsj (top) and \tsj (bottom) for trigger (left panels) and recoil (right panels) jets. The \AuAu plots are shown in the filled markers whereas the open markers represent \ppAA which includes the shaded regions representing systematic uncertainty. The red and black points represent HardCore and Matched jets respectively.}
   \label{fig:twoSubJet}
\end{figure}

Figure~\ref{fig:twoSubJet} shows the subjet \zsj~(top) and \tsj~(bottom) distributions for trigger jets (left panels) with $p_{\rm{T}} > 16$ GeV/$c$ and recoil (right panels) jets with $p_{\rm{T}} > 8$ GeV/$c$. HardCore and Matched jets are shown in the red squares and black circles, respectively, with $R_{\rm{SJ}}=0.1$ subjets. For both the HardCore and Matched jet distributions in each panel of Fig.~\ref{fig:twoSubJet}, the differences between the \AuAu data and the \ppAA reference distribution are negligible indicating, once again, no significant modification of the jet substructure due to jet quenching. Given that the substructure results are similar, we may conclude that the recoil jets in the kinematic range studied in this analysis, fragment in a vacuum-like environment. The $\theta_{\rm{SJ}}$ for Matched jets, in stark contrast to the SoftDrop \rg, peaks at small values which include a natural lower cutoff at the subjet radius. Di-jet pairs are selected based on the recoil Matched jet \tsj, and we define narrow (wide) recoil jets as $0.1 ~ (0.2) < \theta_{\rm{SJ}} < 0.2 ~ (0.3)$. The narrow (wide) jets have a \tsj~finding purity of 98\% (75\%) determined from the correlation matrix shown in the bottom left panel of Fig.~\ref{fig:zsjtsjEmbeffect}.

\section{Di-jet asymmetry dependence on angular scale}
\label{sec:diffAj}

\begin{figure}[h] %  figure placement: here, top, bottom, or page
   \centering
   \includegraphics[trim=110 110 0 10,clip,width=0.8\textwidth]{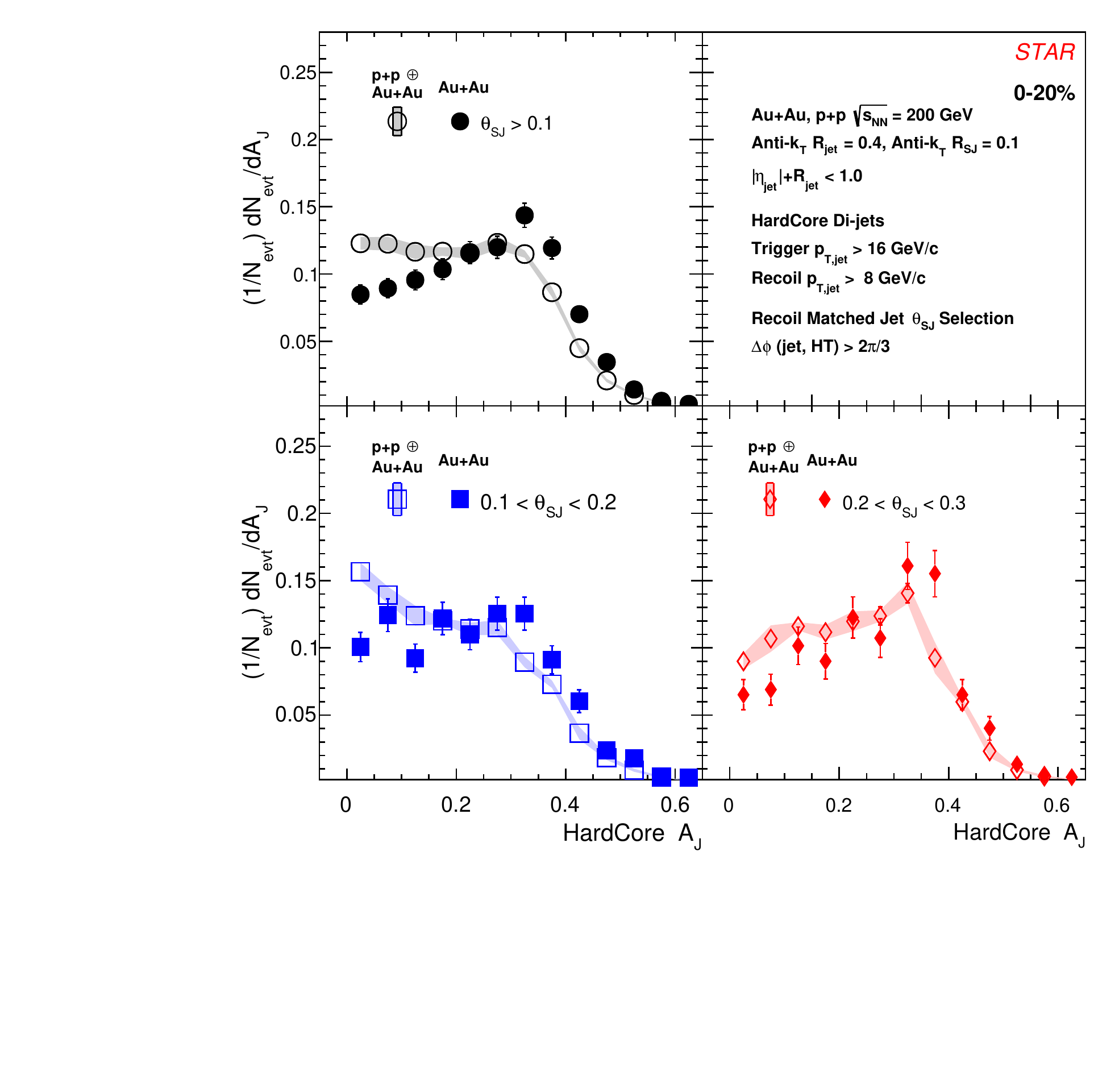}
   \caption{Distributions of HardCore di-jet Asymmetry (\Aj) for events with leading and subleading cuts of 16 GeV/$c$ and 8 GeV/$c$, respectively. The filled markers correspond to \AuAu data and the open markers along with the shaded regions show \ppAA~reference. The black distributions do not have any cuts on \tsj~whereas for the blue and red distributions, selections on recoil Matched jets with $0.1 < \theta_{\rm{SJ}}  < 0.2$ and $0.2 < \theta_{\rm{SJ}} < 0.3$ are applied. The vertical error bars represent statistical uncertainty on the data points.}
   \label{fig:hcaj}
\end{figure}

\begin{figure}[h] %  figure placement: here, top, bottom, or page
   \centering
   \includegraphics[trim=110 110 0 10,clip,width=0.8\textwidth]{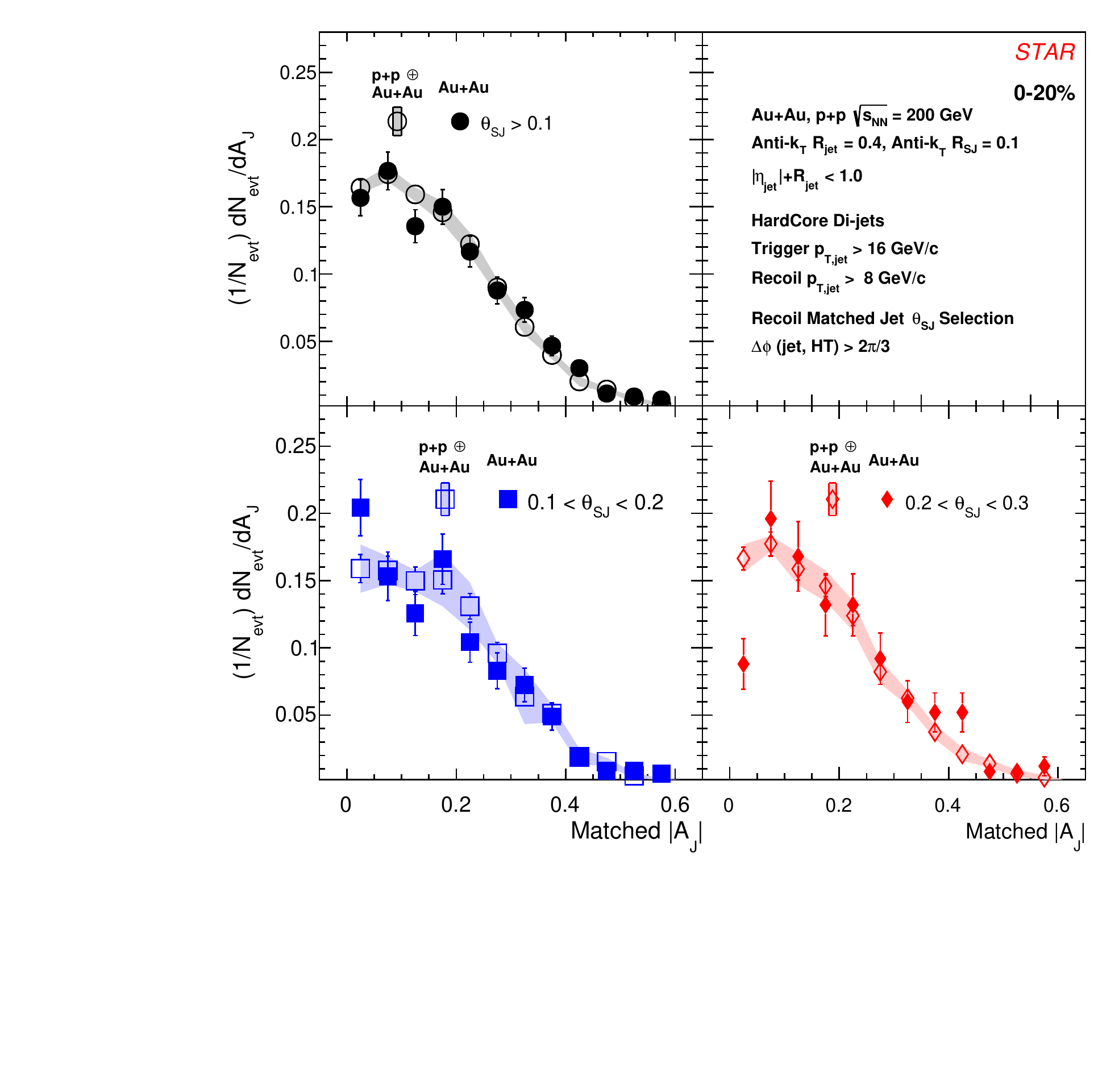}
   \caption{Distributions of Matched di-jet Asymmetry ($|A_{\rm{J}}|$) for events with leading and subleading cuts of $16$ GeV/$c$ and $8$ GeV/$c$ (for HardCore jets), respectively. The filled markers correspond to \AuAu data and the open markers along with the shaded regions show \ppAA reference. The black distributions do not have any cuts on the subjet \tsj~whereas for the blue and red distributions, selections on recoil Matched jets with $0.1 < \theta_{\rm{SJ}}  < 0.2$ and $0.2 < \theta_{\rm{SJ}} < 0.3$ are applied. The vertical error bars represent statistical uncertainty on the data points.}
   \label{fig:mataj}
\end{figure}

The differential measurements of momentum asymmetry for HardCore and Matched di-jets (in $|A_{\rm{J}}|$) are shown in Fig.~\ref{fig:hcaj} and Fig.~\ref{fig:mataj}, respectively. The black, blue and red markers represent recoil jets with selections on their corresponding Matched jet \tsj~with the ranges $[0.1, 0.4], [0.1, 0.2]$ and $[0.2, 0.3]$ for inclusive, narrow and wide jets, respectively. The selection on the \tsj~reflects the available resolution due to the statistics in the data sample. We observe a clear di-jet imbalance indicating jet quenching effects in \AuAu collisions for all HardCore jets including the wide angle jets. The Matched jets on the other hand are momentum balanced at RHIC energies, as is evident by the overlap between the \AuAu (filled) and \ppAA (open) markers. This is consistent with our earlier measurements~\cite{Adamczyk:2016fqm}, and agrees with the observation that both wide and narrow angle matched jets are, respectively, balanced with the reference with little change in their overall shapes. This indicates that the lost energy from the HardCore di-jets is recovered in the softer particles ($0.2 < p_{\rm{T}} < 2.0$ GeV/$c$) distributed around the jet cone of $R=0.4$. While one might expect that for wider jets, the energy loss would spread to larger angles, we observe that within the available experimental resolution of \tsj, there is no significant difference in the energy loss signature of jets with $0.1 < \theta_{\rm{SJ}} < 0.3$. 

\section{Conclusions and discussion}
\label{sec:summary}

We present the first differential measurement of partonic energy loss in \AuAu collisions at \sqrtsn $=200$ GeV for jets tagged via their opening angle. The energy loss is quantified with measurements of the momentum asymmetry \Aj~of specially selected di-jet pairs. The differential nature of these measurements involves identifying and selecting jets of a particular topology or substructure, {\textit i.e.}, those that have narrow vs. wide opening angles. Since we compare \AuAu data to an embedded \ppAA reference, we require the substructure observable to be sensitive to the jet kinematics, while simultaneously being insensitive to the heavy-ion underlying event. For SoftDrop observables with $z_{\rm{cut}} = 0.1$ and $\beta = 0$, we find that the groomed jet radius ($R_{\rm{g}}$) is impacted by the fluctuating underlying event and this results in a significant fraction of jets tagged with a fake splitting as shown in Fig.~\ref{fig:zgrgEmbeffect}. This effect arises from background particles which satisfy the grooming criteria, thus complicating the use of $R_{\rm{g}}$, particularly for larger opening angles.

While the fake split fraction might be reduced by limiting the analysis to high-\pt jets or by varying the SoftDrop grooming parameters, we instead introduced a new class of subjet observables, \zsj~and \tsj. In the present analysis the subjets are reclustered via the anti-$k_{\rm{T}}$ algorithm with the jet's constituents as input and with a smaller resolution parameter $R_{\rm{SJ}} = 0.1$. We present the fully corrected subjet observables (\tsj~and \zsj) for \pp collisions, both of which display a gradual change in the jet shape from a broad to a narrow distribution as the jet momentum increases. This evolution with jet momentum, as shown in Fig.~\ref{fig:pptwosubjet}, indicates a transition from large momentum sharing (\zsj~peaked at large values) between the two leading subjets to a gradually more asymmetric momentum sharing for higher \pt jets. More importantly for our study, these subjet observables meet the requirement of being both sensitive to the jet kinematics and insensitive to the heavy-ion background via an absolute \pt threshold on the subjets as shown in the correlation studies in Fig.~\ref{fig:zsjtsjEmbeffect}. This achievement contrasts with the SoftDrop method which uses a fractional momentum cut, thus has a jet \pt dependent bias due to the combination of quenching and surface bias effects on jet selection criteria at RHIC energies.

In comparing the substructure distributions for \AuAu data and the \ppAA reference for both trigger/recoil HardCore/Matched jets, we observe no significant differences in all cases, indicating that the splittings identified in jets via the subjet method are vacuum-like. Measurements of \Aj~for recoil jets of varying \tsj~demonstrate no significant differences in the momentum balance/imbalance of Matched/HardCore di-jet pairs for recoil jets with $0.1 < \tsj < 0.2$ or $0.2 < \tsj < 0.3$. These results support the conclusion that these particular selected di-jets do not undergo significantly different jet-medium interactions under varying angular scales.

We can now develop a consistent picture of partonic energy loss for specially selected di-jets at RHIC energies based on three significant features that we observe in our data. The first is that these recoil jets are expected to have smaller path-lengths in the medium on average, owing to the restrictive di-jet requirements which favor tangential production vertices, in comparison to an inclusive or semi-inclusive jet population. The second is the observation that the jet substructure distributions are comparable for \AuAu and \ppAA indicating vacuum like splitting. Third, the recovery of the quenched energy for recoil Matched jets is independent of the jet opening angle measured via the \tsj. Thus, we infer that the recoil jet's first hard splitting during jet evolution possibly happens at formation times comparable to the shorter in-medium path length for tangential di-jets, resulting in vacuum like distributions. Given that the HardCore recoil jets do undergo quenching, as shown by an imbalanced \Aj, the medium interaction that these jets undergo happens at earlier times when the hard-scattered parton is traversing the medium.

These three features, together with the surface bias of unmodified trigger jets, lead us to a qualitative interpretation of the data that energy loss in these recoil jets is due to medium induced radiation from a single color charge. Because of the relatively small-scale resolution of the subjet opening angle in this measurement $(\Delta \theta_{\rm{SJ}} = 0.1)$, we were able to observe di-jet balance/imbalance for both narrow and wide jets. From the similarity of the results for narrow and wide jets we conclude that there is no observational evidence of the characteristic signature of coherent or de-coherent energy loss as the range of sampled jet opening angles encompasses the medium coherence length scale. 
The differential measurements presented here can now be utilized in stringent tests of various quenching models, and also interpretations resulting from jet selection and fragmentation biases. These studies lead the community towards a study of soft gluon radiation in the QGP, as in the QCD analog of the Landau-Pomeranchuk-Migdal effect~\cite{Landau:1953gr, Migdal:1956tc, Wang:1994fx, Zapp:2011ya}, which has long been expected to be a significant factor in parton energy loss at RHIC.

\section*{Acknowledgements}

We thank the RHIC Operations Group and RCF at BNL, the NERSC Center at LBNL, and the Open Science Grid consortium for providing resources and support.  This work was supported in part by the Office of Nuclear Physics within the U.S. DOE Office of Science, the U.S. National Science Foundation, the Ministry of Education and Science of the Russian Federation, National Natural Science Foundation of China, Chinese Academy of Science, the Ministry of Science and Technology of China and the Chinese Ministry of Education, the Higher Education Sprout Project by Ministry of Education at NCKU, the National Research Foundation of Korea, Czech Science Foundation and Ministry of Education, Youth and Sports of the Czech Republic, Hungarian National Research, Development and Innovation Office, New National Excellency Programme of the Hungarian Ministry of Human Capacities, Department of Atomic Energy and Department of Science and Technology of the Government of India, the National Science Centre of Poland, the Ministry  of Science, Education and Sports of the Republic of Croatia, RosAtom of Russia and German Bundesministerium f\"ur Bildung, Wissenschaft, Forschung and Technologie (BMBF), Helmholtz Association, Ministry of Education, Culture, Sports, Science, and Technology (MEXT) and Japan Society for the Promotion of Science (JSPS).

\appendix*
\section{Detector effects and comparisons}
\label{sec:appA}

\begin{figure}[h] %  figure placement: here, top, bottom, or page
   \centering
   \includegraphics[width=0.6\textwidth]{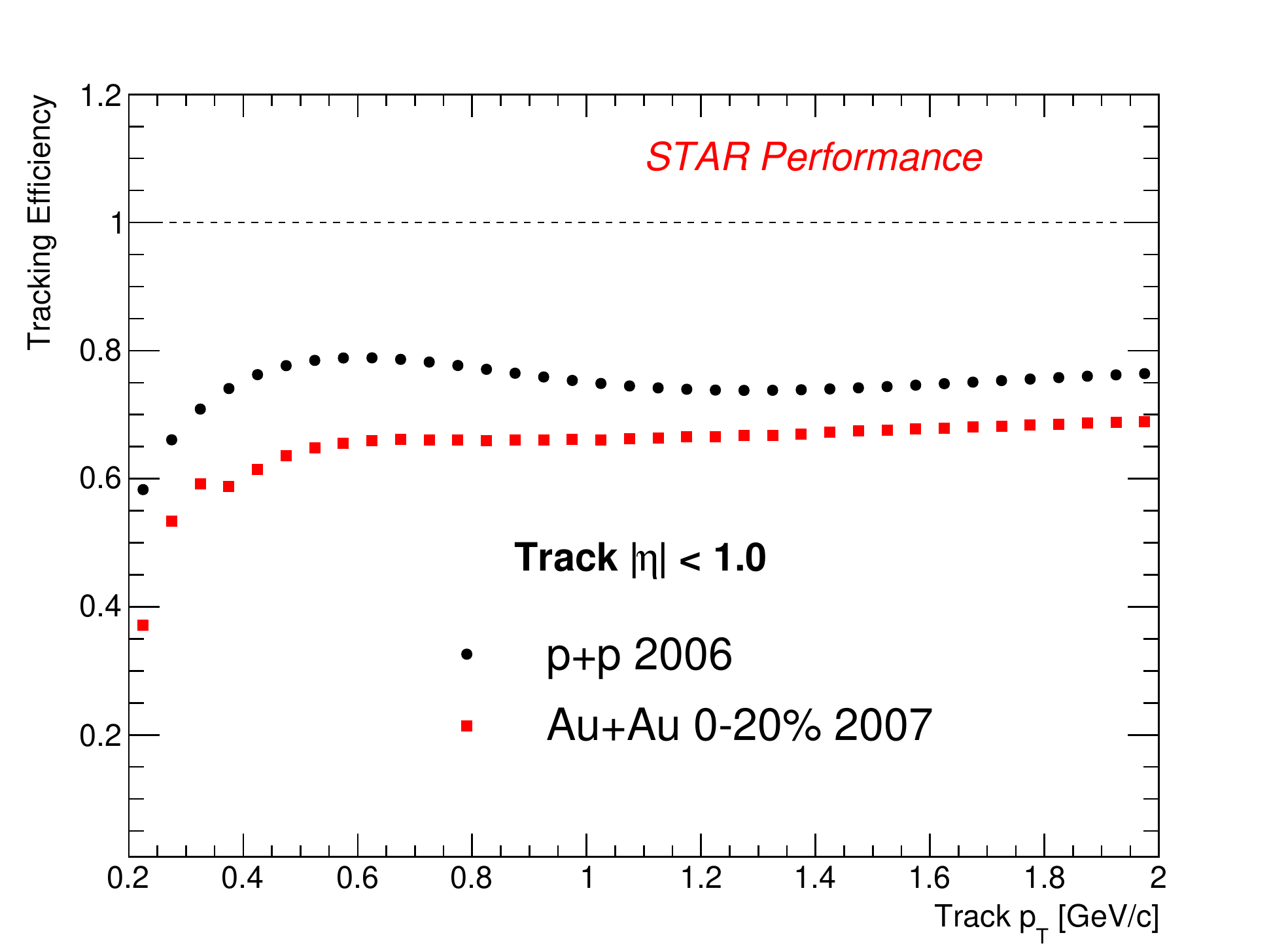}
   \caption{TPC tracking efficiencies for the 2006 \pp and 2007 \AuAu datasets utilized in the embedding studies for tracks within $|\eta| < 1.0$.}
   \label{fig:trackeffcomp}
\end{figure}

The \AuAu data in this publication are compared to an embedded reference at the detector-level which presents the measurement without any correction for detector effects. 
We therefore provide the relevant performance parameters for the STAR detector, mainly the TPC and the BEMC. This enables predictions of 
MC models or theoretical calculations to be directly applied to the detector-level data.
For charged-particle tracks in the TPC, 
the tracking efficiency is shown in Fig.~\ref{fig:trackeffcomp} as a function of the track 
$p_{\rm{T}}$ for particles at mid-rapidity ($|\eta| < 1.0$). The red and black markers show 
the efficiencies for \pp and \AuAu 0-20\% events taken during 2006 and 2007, respectively. 
The tracking efficiency is also assumed to be flat as a function of track momentum for 
$2.0 < p_{\rm{T}} < 30$ GeV/$c$ for both datasets.
The TPC also produces a momentum smearing which is modeled by

\begin{equation}
    \sigma = -0.026 + 0.02 \cdot p^{\rm{true}}_{\rm{T}} + 0.003 \cdot (p^{\rm{true}}_{\rm{T}})^{2},    
\end{equation}
taken to be the same for both \pp and \AuAu collisions.

The BEMC has a spatial segmentation of $0.05 \times 0.05$ in ($\eta,\phi$) with an energy resolution of $\sigma (E_{\rm{T}}) = 14\%/\sqrt{E_{\rm{T}}}$~\cite{Beddo:2002zx}. The hadronic correction procedure described at the beginning of Sect.~\ref{sec:dataset} ensures that the energy deposited by charged-particles in the BEMC is not double-counted, such that $\sigma (E_{\rm{T}})$ estimates the error in the neutral energy of a jet.  

In addition to the preceding detector effects, the impact of the heavy-ion underlying event on the jet momentum and substructure observables should be taken into account for direct comparison with the data presented here. These effects for the HardCore and Matched jet momenta are presented in the supplemental material of an earlier publication~\cite{Adamczyk:2016fqm}. The left panels of Fig.~\ref{fig:zgrgEmbeffect} and Fig.~\ref{fig:zsjtsjEmbeffect} of this reference show the effect of the heavy-ion underlying event on the substructure observables.  

\bibliography{auaurun7_substructure_prc_paper_raghav_2021.bib}

\end{document}